\documentclass[prb,preprint,showpacs,showkeys,preprintnumbers,amsmath,amssymb]{revtex4-1}

\usepackage{amsmath,amssymb,bm,epsfig}
\usepackage{dcolumn}
\usepackage{bm}
\usepackage{feynmf}

\begin{document}


\title{Quantum and thermal fluctuations of a thin elastic plate}

\author{Dennis P. Clougherty$^{1, 2}$ and Eliot Heinrich$^1$}

\affiliation{
$^1$ Department of Physics, 
University of Vermont, 
Burlington, VT 05405-0125}

\affiliation{
$^2$ JILA, National Institute of Standards and Technology and University of Colorado, 440 UCB, Boulder, CO 80309}

\date{\today}

\begin{abstract}
We consider a Hamiltonian description of the vibrations of a clamped, elastic circular plate. The Hamiltonian of this system features a potential energy with two distinct contributions: one that depends on the local mean curvature of the plate, and a second that depends on its Gaussian curvature. We quantize this model using a complete, orthonormal set of eigenfunctions for the clamped, vibrating plate. The resulting quanta are the flexural phonons of the thin circular plate. As an application, we use this quantized description to calculate the fluctuations in displacement of the plate for arbitrary temperature.  We compare the fluctuation profile with that from an elastic membrane under tension.  At low temperature, we find that while both profiles have a circular ring of local maxima, the ring in the membrane profile is much more pronounced and sharper.  We also note that with increasing temperature the plate profile develops two additional rings of extrema.

\end{abstract}

%

\maketitle
\section{Introduction}

Vibrating mesoscopic beams, membranes, and plates are now frequently used as elements of so-called ``hybrid'' quantum devices and precision sensors.   In an effort to understand these devices in detail, it is worthwhile to develop a fully quantum mechanical model of the dynamics of these elements.  In addition to aiding in the development of new quantum technologies,  such descriptions of the quantum dynamics of mesoscopic elastic solids might be used to conceive of fundamental tests of quantum theory that create, entangle, and control nonclassical mechanical states of the solid \cite{optomechanics,entangle2018}.

Experimental studies of suspended graphene samples \cite{mceuen08} prompted prior theoretical work \cite{DPC2013} to quantize the vibrations of a clamped elastic membrane under tension.   These results were subsequently applied in studies of quantum sticking of ultracold atoms to suspended 2D materials \cite{dpc11,dpcSPR2017,Sengupta2018}.

Recent experimental work has demonstrated that it is now possible to probe Fock states of mechanical ``quantum drums''\cite{lehnert2018} and acoustic resonators \cite{Schoelkopf2018}.  While the vibrations of single-layer and multilayer graphene have been previously analyzed using continuum elasticity theory \cite{[][{.   (We note an exponent of 2 in the bending energy density is missing in Eq.~2.)}] deandres2012}, such a classical description is insufficient to describe low temperature systems with a small, definite number of flexural phonons where quantization is important.  In addition to multilayer 2D materials, the results obtained in this work on the quantization of the flexural modes of thin elastic plates will have application to low temperature optomechanical systems that use, for example,  thin films of silicon nitride \cite{microdisk2006, regal2009, regal2012, bonaldi2016, painter2016}.   

We begin by employing  canonical quantization methods to construct a quantum mechanical description of a thin, circular, vibrating elastic plate subject to clamped boundary conditions. As an application of this quantum description, we calculate both the zero-point and thermal fluctuations of the plate, and we compare these results with those from a 2D elastic membrane under tension, a model previously used to describe  the vibrational dynamics of suspended single-layer graphene.

\section{Lagrangian}

\begin{figure}
\includegraphics[width=9cm]{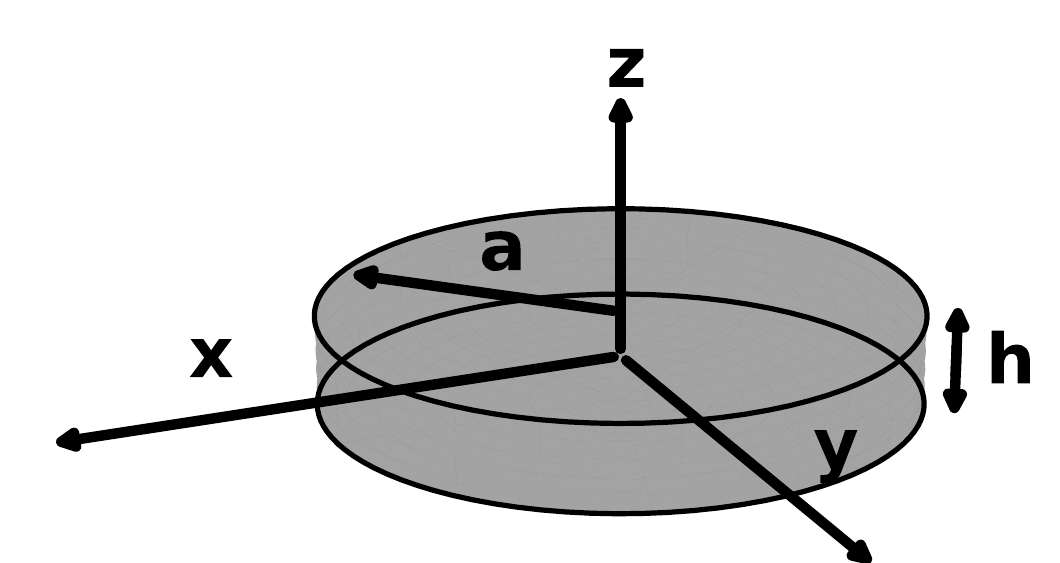}
\caption{\label{fig:plate} Sketch of an elastic plate (radius $a$, thickness $h$) in equilibrium.}
\end{figure}

We start by summarizing the Lagrangian description of the classical flexural vibrations of a thin elastic plate. Let $u({r}, \theta,t)$ be the elastic displacement field normal to the plate in equilibrium (see Fig.~\ref{fig:plate}).  The Lagrange density \cite{landau} is given by 
\begin{equation}
    \mathcal{L} = \mathcal{T} - \mathcal{U} = \frac{1}{2}\rho h \left(\frac{\partial u}{\partial t}\right)^2 - \frac{D}{2}\left(\nabla^2 u\right)^2 - D(1-\nu)\left[\left(\frac{\partial^2 u}{\partial x \partial y}\right)^2 - \frac{\partial^2 u}{\partial x^2}\frac{\partial^2 u}{\partial y^2}\right]
    \label{lagrangian}
\end{equation}
where  $\rho$ is the volume mass density, $\nu$ is the Poisson ratio, and the flexural rigidity $D$ is related to the Young's modulus $E$ by $D={E h^3\over 12 (1-\nu^2)}$.  
For a thin plate, the displacement is uniform along the plate thickness $h$.  Hence, $\nabla^2$ here is the 2D Laplacian.


We note that the strain energy density contains two contributions: one that depends on the mean curvature of the plate $\nabla^2 u$, and a second proportional to its Gaussian curvature $K=(u_{xy}^2 -  u_{xx} u_{yy})$.  This is in contrast to the strain energy density of an elastic membrane under tension which depends on the square of the magnitude of the displacement gradient \cite{DPC2013} $|\nabla u|^2$. 

The equation of motion follows directly from this Lagrange density
\begin{equation}
    \frac{\partial^2 u}{\partial t^2} + \frac{D}{\sigma}\nabla^4 u = 0
    \label{eom}
\end{equation}
where $\sigma=\rho h$ and $\nabla^4\equiv \nabla^2 \nabla^2$.

For time-harmonic solutions of the form
\begin{equation}
    u({\bf r},t) = w(r,\theta) e^{-i \omega t}
\end{equation}
Eq.~\ref{eom} becomes
\begin{equation}
    (\nabla^4 - k^4)w({r},\theta) = 0
    \label{modes}
\end{equation}
where $k^4 = \frac{\omega^2 \sigma}{D}$.
For a clamped plate, the following boundary conditions are imposed:  $w(a,\theta) = \frac{\partial w(r,\theta)}{\partial r}|_{r=a}= 0$ where $a$ is the radius of the circular plate. 

Regularity at the plate's center leaves a solution to Eq.~\ref{modes} of the following form
\begin{equation}
w_m(r,\theta)=(A_m J_m(k r)+B_m I_m(k r)) \exp(i m \theta)
\end{equation}
The boundary conditions yield the additional constraint on allowable values of $k$; namely, 
\begin{equation}\label{eqn:constraint}
    I_m(k a) J_{m+1}(k a) + J_m(k a) I_{m+1}(k a) = 0
\end{equation}
A selected set of the lowest eigenvalues satisfying this root condition is given in Table~\ref{table:kmn}.

\begin{table}
\caption{\label{table:kmn} Lowest eigenvalues $k_{mn}a$ for root condition of Eq.~\ref{eqn:constraint}.}
\begin{ruledtabular}
\begin{tabular}{c|cccccc}
n & m=0 & m=1&m=2&m=3&m=4&\\
\hline
1 & 3.196&  4.611&5.906&7.144&8.347&\\ 
2 & 6.306&  7.799&9.197&10.536&11.837&\\
3 &9.439& 10.958&12.402&13.795&15.150&\\
4&12.577&14.109&15.579&17.005&18.396&\\

\end{tabular}
\end{ruledtabular}
\end{table}

The  flexural modes of the thin plate are thus given by
\begin{eqnarray}
w_{mn}({\bf r}) &=& R_{mn}(r) e^{i m \theta}\nonumber \\
 &=& \mathcal{N}_{mn}\left[J_{m}(k_{mn}r) -\bigg({J_m(k_{mn} a)\over I_m(k_{mn} a)}\bigg) I_m (k_{mn}r)\right] e^{i m \theta}
 \label{eigenmodes}
\end{eqnarray}
where $k_{mn}$ is a solution to the root condition of Eq.~\ref{eqn:constraint}.

\begin{figure}
\includegraphics[width=12cm]{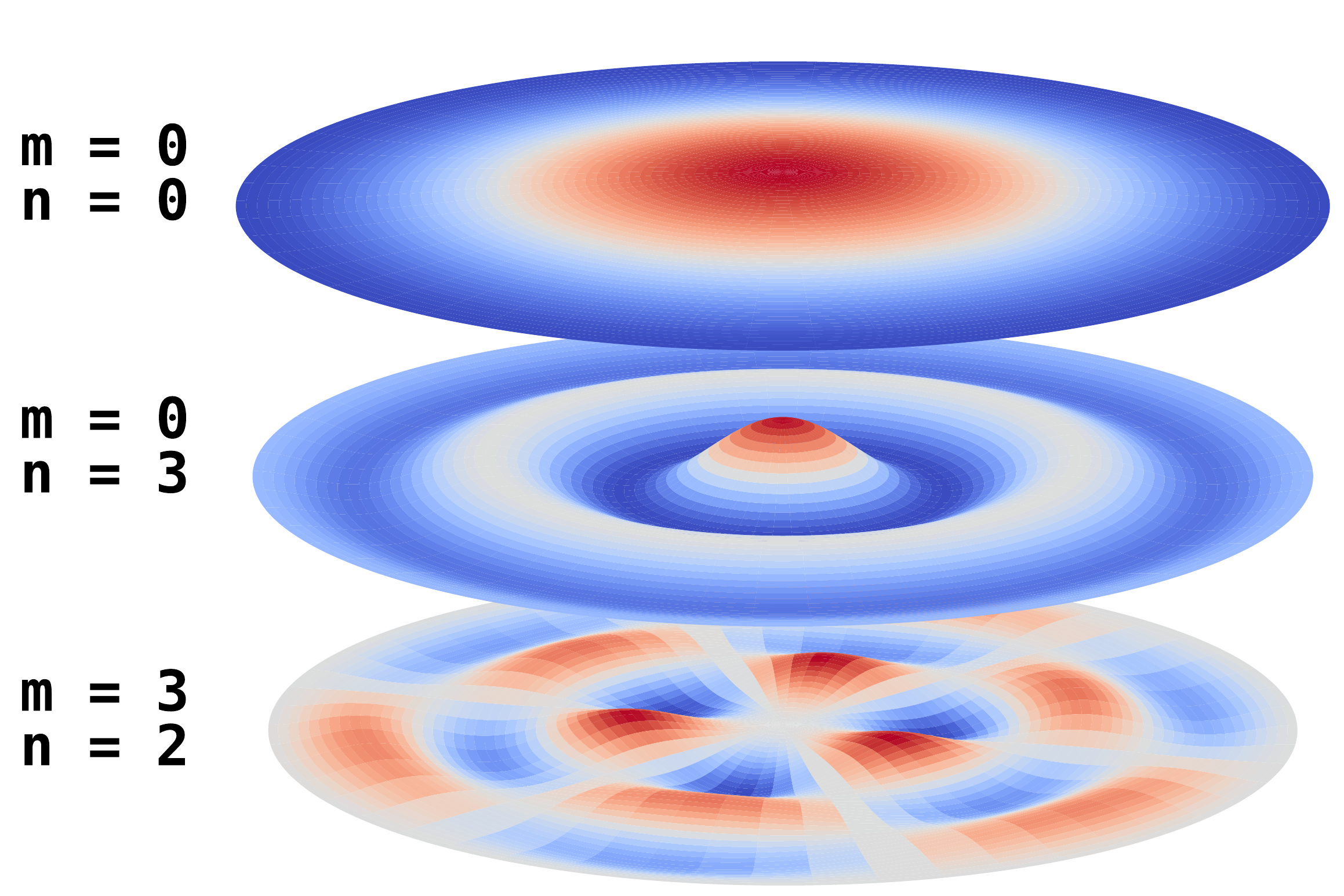}
\caption{\label{fig:modes} Plot of selected low frequency eigenmodes $w_{mn}({\bf r})$ for the vibrating plate.}
\end{figure}

We choose the normalization of the normal mode solutions $w_{mn}$ so that  
\begin{equation}
 \int w^*_{mn}({\bf r}) w_{m'n'}({\bf r}) d^2r = \delta_{mm'}\delta_{nn'}
\end{equation}
The following normalization constant results
\begin{equation}
    \mathcal{N}_{mn} = \sqrt{\frac{1}{\pi a^2} \frac{I_{m}(k_{mn}a)}{J_{m}(k_{mn}a) f(k_{mn} a)}} 
    \label{normal}
\end{equation}
where $f(\lambda)={{2J_{m}(\lambda)I_{m}(\lambda) + J_{m-1}(\lambda)I_{m+1}(\lambda) + J_{m+1}(\lambda)I_{m-1}(\lambda)}}$.

It is worth noting that for circularly symmetric solutions ($m=0$), the normalization constant simplifies considerably. With the use of Bessel function recursion relations, we obtain
\begin{equation}
    \mathcal{N}_{0n} = \sqrt{\frac{1}{2 \pi a^2}} \frac{1}{|J_{0}(k_{0n}a)| } 
    \label{m0}
\end{equation}

\section{Hamiltonian}
We now turn to the Hamiltonian density which is constructed from the Lagrange density of Eq.~\ref{lagrangian}. We obtain that
\begin{equation}
    \mathcal{H} = \frac{1}{2 \sigma}\Pi^2 + \frac{D}{2}\left(\nabla^2 u\right)^2 + D(1-\nu)\left[\bigg({\partial\over\partial r}\bigg({1\over r}{\partial u\over \partial\theta}\bigg)\bigg)^2-{\partial^2 u\over\partial r^2}\bigg({1\over r}{\partial u\over\partial r}+{1\over r^2}{\partial^2 u\over\partial\theta^2}\bigg)\right]
\end{equation}
where the canonical momentum density is $\Pi({\bf r},t) = \frac{\partial \mathcal{L}}{\partial \dot{u}}$.  (We have rewritten the Gaussian curvature contribution to the strain energy density of the plate in polar coordinates.)

The displacement field $u$  can be expanded in normal modes.  Thus, 
\begin{equation}
u=\sum_{m=-\infty}^\infty \sum_{n=1}^\infty Q_{mn} w_{mn}
\end{equation}

Using the 2D divergence theorem, it can be shown that 
\begin{equation}
\int (\nabla^2 u)^2 d^2 r=\int u \nabla^2 \nabla^2 u\ d^2 r
\end{equation}
Hence, using the equation of motion (Eq.~\ref{eom}) and orthonormality of the normal modes, we obtain that 
\begin{equation}
\int (\nabla^2 u)^2 d^2 r= \sum_{m=-\infty}^\infty \sum_{n=1}^\infty Q_{mn} Q_{{\bar m}n} k^4_{mn}
\end{equation}
(We use notation where ${\bar m}\equiv -m$.)

Expanding the momentum density $\Pi$ in normal modes
\begin{equation}
\Pi=\sum_{m=-\infty}^\infty \sum_{n=1}^\infty P_{{\bar m}n} w_{mn}
\end{equation}
we obtain that the kinetic energy  goes as
\begin{equation}
{1\over 2\sigma}    \int \Pi^2 d^2 r = {1\over 2\sigma} \sum\limits_{m,n} P_{mn} P_{\bar{m}n}
\end{equation}

The total Gaussian curvature contribution to the potential energy will vanish by an application of the Gauss-Bonnet theorem \cite{topology}
\begin{equation}
    \int_M K d^2 r +\int_{\partial M} k_g ds=2\pi \chi_M
\end{equation}
The Euler characteristic $\chi_M$ for the manifold $M$  (a closed circular disk) is 1, and the geodesic curvature of the circular boundary $k_g=1/a$.  Thus, $\int K d^2 r = 0$.

The Hamiltonian  then may be rewritten as
\begin{equation}
    H = \sum\limits_{m,n} \left( \frac{1}{2\sigma} P_{\bar{m}n} P_{mn} + \frac{\sigma \omega^2_{mn} }{2}Q_{\bar{m}n} Q_{mn}\right)
\end{equation}

\section{Canonical quantization}
To formulate a quantum description, we introduce in the usual way \cite{DPC2013}  the creation and annihilation operators for the vibrational modes 
\begin{eqnarray}
b_{mn}&={i}\sqrt{1\over 2\sigma\hbar\omega_{mn}}P_{\bar m n}+\sqrt{\sigma\omega_{mn}\over 2\hbar}  Q_{mn}\cr
b^\dagger_{mn}&=-{i}\sqrt{1\over 2\sigma\hbar\omega_{mn}}P_{mn}+\sqrt{\sigma\omega_{mn}\over 2\hbar} Q_{\bar m n}
\label{creation}
\end{eqnarray}

We impose the canonical commutation relation between displacement and momentum density 
\begin{equation}
[u({\bf r}), \Pi({\bf r'})]=i\hbar \delta({\bf r}-{\bf r'})
\end{equation}
and obtain the following Hamiltonian 
\begin{equation}
H=\sum_{n, m} \hbar\omega_{mn}\bigg(b^\dagger_{mn} b_{mn}+\frac{1}{2}\bigg)
\end{equation}

The displacement field $u({\bf r})$ can then be written in quantized form 
\begin{equation}\label{eqn:uq}
    u({\bf r}) = \sum\limits_{m,n} (U_{mn}({\bf r}) b_{mn} + U^*_{mn}({\bf r}) b^\dag_{{m}n})
\end{equation}
where $U_{mn}({\bf r})= \sqrt{\frac{\hbar}{2\sigma \omega_{mn}}} w_{mn}({\bf r})$.

This result differs from the corresponding result of an elastic membrane under tension in two important ways: (1) we observe that the flexural phonons of the plate have a different (quadratic) dispersion from the linear dispersion of transverse acoustic phonons for the membrane under tension, and (2) the  spatial (radial) character of the normal modes $w_{mn}$ also differs. We will see that these differences lead to contrasting behavior in the pattern of fluctuations of the surface.

\section{Fluctuations of the plate}
\begin{figure}
\includegraphics[width=16cm]{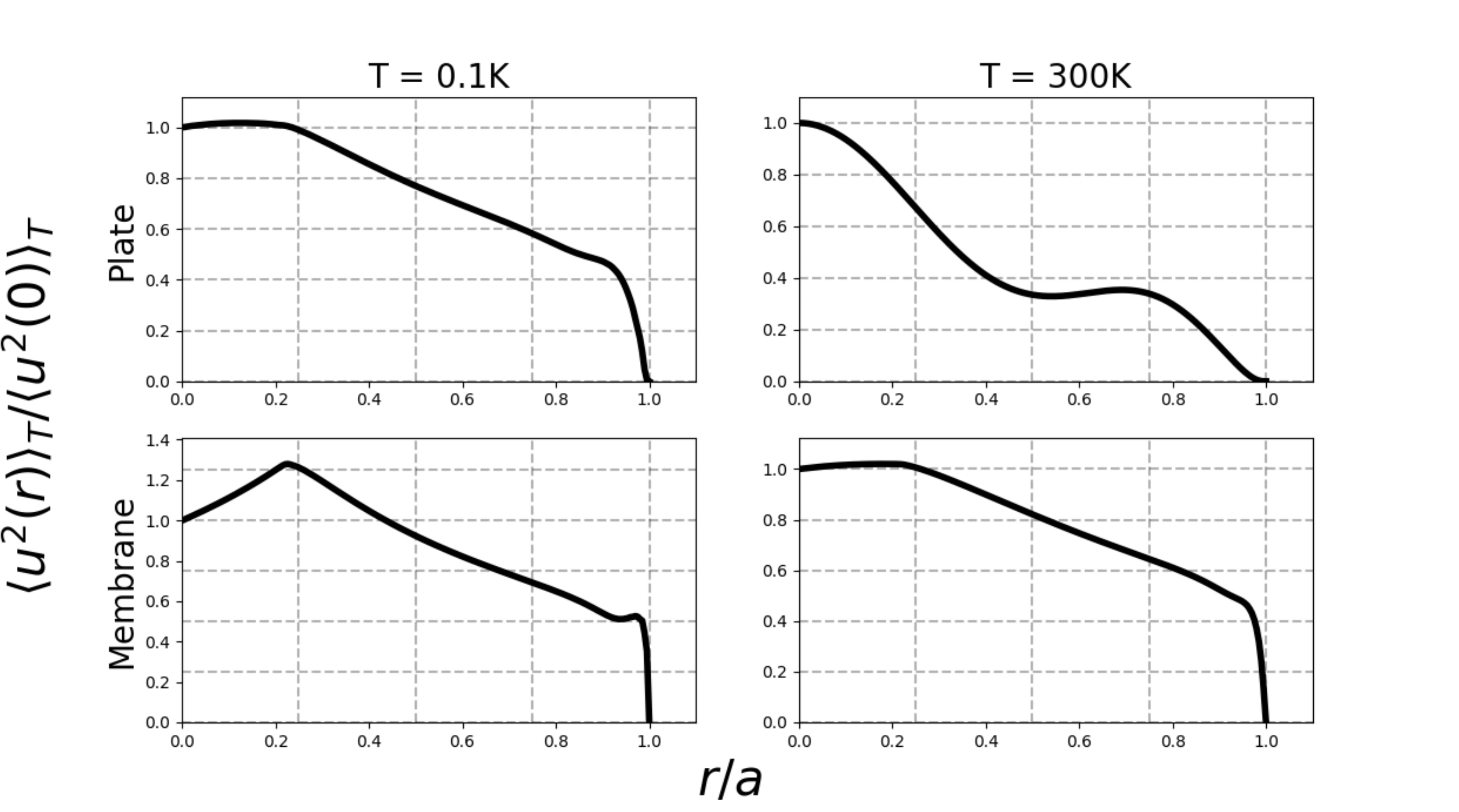}
\caption{\label{fig:radial} Plot of thermal average of the squared displacement (scaled) $ \langle u^2(r)\rangle_T/\langle u^2(0)\rangle_T$ versus (reduced) radius $r/a$ for (top row) a silicon nitride plate with $\Theta_c= 80$ mK (with $\Theta_c\equiv \hbar\omega_c/k_B$) and (bottom row) an elastic membrane under tension. Both have radius of $a=1 \mu$m.}
\end{figure}

As an application of the quantum description of the flexural vibrations of the thin elastic plate, we calculate the fluctuations in the displacement of the plate.  We then compare the result to the corresponding quantity for a 2D membrane.  From Eq. (\ref{eqn:uq}), we obtain the thermal average of the square of the displacement  $\langle u^2({\bf r})\rangle_T $  as
\begin{eqnarray}
    \langle u^2({r})\rangle_T &=& \sum\limits_{m, n} \bigg({\hbar^2\over 2\sigma\omega_{mn}}\bigg) |R_{mn}({r})|^2 (2\langle b^\dagger_{mn}b_{mn}\rangle_T+1)\nonumber\\
    &=&\sum\limits_{m, n} \bigg({\hbar^2\over 2\sigma\omega_{mn}}\bigg) |R_{mn}({r})|^2 \coth({\beta\hbar\omega_{mn}\over 2})
\label{u2}
    \end{eqnarray}
The  function $\langle u^2({\bf r})\rangle_T $ is independent of angle $\theta$ and is a varying  function of distance from the center (see Fig.~\ref{fig:radial}).   We will refer to this function as the fluctuation profile.


For the case of the center of the plate, the thermal average of the square of the displacement  $\langle u^2({0})\rangle_T $ becomes
\begin{equation}
    \langle u^2(0)\rangle_T = \sum\limits_n |U_{0n}({0})|^2 \coth({\beta\hbar\omega_{0n}\over 2})
\label{rms}
    \end{equation}
    We evaluate the sum using the continuum approximation, valid for mesoscopic-sized plates.
    Using a vibrational density of symmetric ($m=0$) flexural modes ${\cal D}(\omega)={a \over 2\pi}\big({\sigma\over D} \big)^{\frac{1}{4}}\omega^{-\frac{1}{2}}$, Eq.~\ref{rms} becomes 
    \begin{equation}
      \langle u^2(0)\rangle_T\approx \int_{\omega_c}^{\omega_D} d\omega {\cal D}(\omega) |U(\omega)|^2 \coth({\beta\hbar\omega\over 2})
    \end{equation}
where  $\omega=\alpha k^2$ with $\alpha=\sqrt{D\over\sigma}$.  The low-frequency cutoff ${\omega_c}$ is a result of the finite size of the plate.  Thus, ${\omega_c}\propto 1/a^2$, a result of the quadratic dispersion of the flexural modes.  We take the high-frequency cutoff $\omega_D$ to be the Debye frequency for the plate.

\begin{table}
\caption{\label{table:compare}Comparison of fluctuations of  displacement $\langle u^2(0)\rangle_T$ for a membrane and a thin plate.}
\begin{ruledtabular}
\begin{tabular}{c|ccc}
System & $T=0$  & $T\gg \Theta_D$&\\
\hline
Membrane & ${\hbar\omega_D\over 4\pi\sigma v_s^2}$&  ${k_B T\over 2\pi\sigma v_s^2}\ln\big({\omega_D\over\omega_c}\big)$&\\ 
Plate & $\sqrt{\hbar^2\over 64\pi^2\sigma D}\ln\big({\omega_D\over\omega_c}\big) $&  ${1\over \sqrt{16\pi^2\sigma D}}\big({k_B T\over\omega_c}\big)$&
\end{tabular}
\end{ruledtabular}
\end{table}

\begin{figure}
\includegraphics[width=16cm]{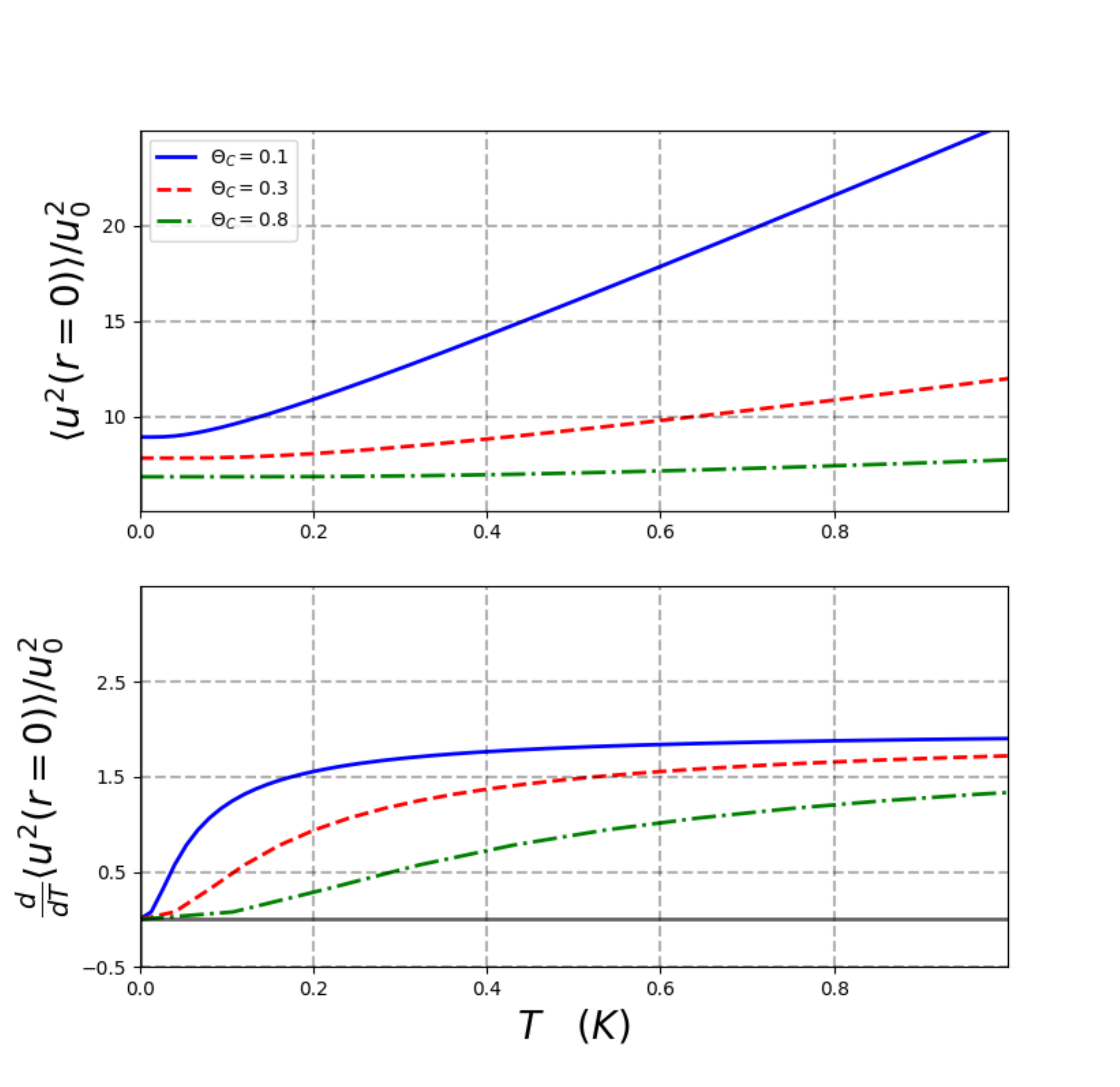}
\caption{\label{fig:u2} Plot of thermal average of the squared displacement of the center of the plate (scaled) $ \langle u^2(0)\rangle_T/u_0^2$ versus  temperature $T$ (K) for a silicon nitride plate (top). Derivative with respect to temperature of the average squared displacement at the center versus temperature $T$ (bottom).   We take $\Theta_D= 850 $ K (Ref.~\onlinecite{sin-debye2013}),  $\Theta_c= 100$ mK (solid), $300$ mK (dashed), and $800$ mK (dot-dashed).  The zero-point fluctuation scale is set by $u_0^2\equiv\sqrt{\hbar^2\over 64 \pi^2\sigma D}$. }
\end{figure}

We consider $U_{0n}(0)$ in the continuum approximation to obtain $U(\omega)$.  The root condition Eq.~\ref{eqn:constraint}  for large $ka$ has the asymptotic form $\tan(k a -\pi/4)=-1$.  Thus, $|J_0(ka)|\to \sqrt{1\over \pi k a}$ in the continuum approximation.  Using the asymptotic form for $J_m$ and $I_m$, we find that $U$ is given by
\begin{equation}
U(\omega)=\bigg( {\hbar^2 \over 16 a^2\sigma^2\alpha \omega}\bigg)^{\frac{1}{4}}
\label{U}
\end{equation}
Thus, $\langle u^2(0)\rangle_T $ becomes
    \begin{eqnarray}
      \langle u^2(0)\rangle_T &\approx&  \sqrt{\hbar^2\over 64 \pi^2\sigma D}  \int_{\omega_c}^{\omega_D} {d\omega\over \omega}\coth({\beta\hbar\omega\over 2})\nonumber\\
      &\approx&  \sqrt{\hbar^2\over 16 \pi^2\sigma D} \bigg({k_B T\over \hbar\omega_c}\bigg), \ \ T\gg\Theta_D
    \end{eqnarray}
where $\Theta_D$ is the Debye temperature of the plate.

For a 2D membrane under tension \cite{DPC2013}, $U=\sqrt{\hbar\over 4\sigma v_s a}$ and the density of 
symmetric modes is ${\cal D}={a\over \pi v_s}$ where $v_s$ is the transverse speed of sound of the membrane.  This yields an average square displacement of the center given by
    \begin{eqnarray}
      \langle u^2(0)\rangle_T &\approx&    \int_{\omega_c}^{\omega_D} {d\omega}  {\hbar\over 4\pi\sigma v_s^2 }  \coth({\beta\hbar\omega\over 2})\nonumber\\
      &\approx&  {k_B T\over 2\pi\sigma v_s^2 } \ln\bigg({\omega_D\over\omega_c}\bigg),  \ \ T\gg\Theta_D
    \end{eqnarray}
The fluctuations of the plate's center grow inversely with the low-frequency cutoff $\omega_c$, while the fluctuations of a 2D membrane grow logarithmically with $\omega_c$.  At low temperatures such that $k_B T\ll \hbar\omega_c$, the membrane fluctuations approach a finite limit as $\omega_c\to 0$.  In contrast, the plate fluctuations at low temperature diverge logarithmically as $\omega_c\to 0$.
We summarize these results in Table~\ref{table:compare} and Fig.~\ref{fig:u2}.

We observe from Fig.~\ref{fig:radial} that the fluctuation profiles have their maximum value away from the center. A power series expansion of Eq.~\ref{u2} about the center reveals that this is a result of contributions to the displacement fluctuations from $m\ne 0$ modes.  The low temperature membrane profile in Fig.~\ref{fig:radial} clearly shows this  maximum near $r\approx 0.2 a$.  With rotational invariance of the fluctuation profile, this maximum away from the center forms a ring of extrema.

In summary, we have constructed a quantum description of the mechanical vibrations of a thin mesoscopic plate.  Using  canonical quantization, we have obtained a description of the displacement field in terms of flexural phonons.  As an application of quantization, we have calculated the thermal expectation value of the square of the displacement field for a circular sample of suspended silicon nitride with radius of 1 $\mu$m.  

We compare the shape of the fluctuation profile with that of an elastic membrane under tension.  We observe sharper features in the fluctuation profile of the membrane in comparison to the profile of the plate.  We attribute this difference to the difference in form of the strain energy density.  For the plate, this energy density depends on the square of the mean curvature of the displacement field.  This tends to smooth out, for example, the sharp local maximum present in the membrane whose strain energy density varies as the square magnitude of the gradient of displacement.   
We also note that the plate tends to develop a ring of local minima in the fluctuation profile at $r\approx a/2$ with increasing temperature (see Fig.~\ref{fig:radial} top row, right panel), a feature absent in the membrane profile.

Support of this work under NASA grant number 80NSSC19M0143 is gratefully acknowledged.  DPC also acknowledges the gracious hospitality of JILA and the partial support of a JILA Visiting Fellowship.

\bibliographystyle{apsrev4-1}
\bibliography{adp}

\begin{thebibliography}{18}%
\makeatletter
\providecommand \@ifxundefined [1]{%
 \@ifx{#1\undefined}
}%
\providecommand \@ifnum [1]{%
 \ifnum #1\expandafter \@firstoftwo
 \else \expandafter \@secondoftwo
 \fi
}%
\providecommand \@ifx [1]{%
 \ifx #1\expandafter \@firstoftwo
 \else \expandafter \@secondoftwo
 \fi
}%
\providecommand \natexlab [1]{#1}%
\providecommand \enquote  [1]{``#1''}%
\providecommand \bibnamefont  [1]{#1}%
\providecommand \bibfnamefont [1]{#1}%
\providecommand \citenamefont [1]{#1}%
\providecommand \href@noop [0]{\@secondoftwo}%
\providecommand \href [0]{\begingroup \@sanitize@url \@href}%
\providecommand \@href[1]{\@@startlink{#1}\@@href}%
\providecommand \@@href[1]{\endgroup#1\@@endlink}%
\providecommand \@sanitize@url [0]{\catcode `\\12\catcode `\$12\catcode
  `\&12\catcode `\#12\catcode `\^12\catcode `\_12\catcode `\%12\relax}%
\providecommand \@@startlink[1]{}%
\providecommand \@@endlink[0]{}%
\providecommand \url  [0]{\begingroup\@sanitize@url \@url }%
\providecommand \@url [1]{\endgroup\@href {#1}{\urlprefix }}%
\providecommand \urlprefix  [0]{URL }%
\providecommand \Eprint [0]{\href }%
\providecommand \doibase [0]{http://dx.doi.org/}%
\providecommand \selectlanguage [0]{\@gobble}%
\providecommand \bibinfo  [0]{\@secondoftwo}%
\providecommand \bibfield  [0]{\@secondoftwo}%
\providecommand \translation [1]{[#1]}%
\providecommand \BibitemOpen [0]{}%
\providecommand \bibitemStop [0]{}%
\providecommand \bibitemNoStop [0]{.\EOS\space}%
\providecommand \EOS [0]{\spacefactor3000\relax}%
\providecommand \BibitemShut  [1]{\csname bibitem#1\endcsname}%
\let\auto@bib@innerbib\@empty
\bibitem [{\citenamefont {Aspelmeyer}\ \emph {et~al.}(2014)\citenamefont
  {Aspelmeyer}, \citenamefont {Kippenberg},\ and\ \citenamefont
  {Marquardt}}]{optomechanics}%
  \BibitemOpen
  \bibfield  {author} {\bibinfo {author} {\bibfnamefont {M.}~\bibnamefont
  {Aspelmeyer}}, \bibinfo {author} {\bibfnamefont {T.~J.}\ \bibnamefont
  {Kippenberg}}, \ and\ \bibinfo {author} {\bibfnamefont {F.}~\bibnamefont
  {Marquardt}},\ }\href@noop {} {\emph {\bibinfo {title} {Cavity
  Optomechanics}}}\ (\bibinfo  {publisher} {Springer-Verlag},\ \bibinfo {year}
  {2014})\BibitemShut {NoStop}%
\bibitem [{\citenamefont {Ockeloen-Korppi}\ \emph {et~al.}(2018)\citenamefont
  {Ockeloen-Korppi}, \citenamefont {DamskŠgg}, \citenamefont {Pirkkalainen},
  \citenamefont {Asjad}, \citenamefont {Clerk}, \citenamefont {Massel},
  \citenamefont {Woolley},\ and\ \citenamefont {Sillanp\"a\"a}}]{entangle2018}%
  \BibitemOpen
  \bibfield  {author} {\bibinfo {author} {\bibfnamefont {C.~F.}\ \bibnamefont
  {Ockeloen-Korppi}}, \bibinfo {author} {\bibfnamefont {E.}~\bibnamefont
  {DamskŠgg}}, \bibinfo {author} {\bibfnamefont {J.-M.}\ \bibnamefont
  {Pirkkalainen}}, \bibinfo {author} {\bibfnamefont {M.}~\bibnamefont {Asjad}},
  \bibinfo {author} {\bibfnamefont {A.~A.}\ \bibnamefont {Clerk}}, \bibinfo
  {author} {\bibfnamefont {F.}~\bibnamefont {Massel}}, \bibinfo {author}
  {\bibfnamefont {M.~J.}\ \bibnamefont {Woolley}}, \ and\ \bibinfo {author}
  {\bibfnamefont {M.~A.}\ \bibnamefont {Sillanp\"a\"a}},\ }\href@noop {}
  {\bibfield  {journal} {\bibinfo  {journal} {Nature}\ }\textbf {\bibinfo
  {volume} {556}},\ \bibinfo {pages} {478Ð482} (\bibinfo {year}
  {2018})}\BibitemShut {NoStop}%
\bibitem [{\citenamefont {Bunch}\ \emph {et~al.}(2008)\citenamefont {Bunch},
  \citenamefont {Verbridge}, \citenamefont {Alden}, \citenamefont {van~der
  Zande}, \citenamefont {Parpia}, \citenamefont {Craighead},\ and\
  \citenamefont {McEuen}}]{mceuen08}%
  \BibitemOpen
  \bibfield  {author} {\bibinfo {author} {\bibfnamefont {J.~S.}\ \bibnamefont
  {Bunch}}, \bibinfo {author} {\bibfnamefont {S.~S.}\ \bibnamefont
  {Verbridge}}, \bibinfo {author} {\bibfnamefont {J.~S.}\ \bibnamefont
  {Alden}}, \bibinfo {author} {\bibfnamefont {A.~M.}\ \bibnamefont {van~der
  Zande}}, \bibinfo {author} {\bibfnamefont {J.~M.}\ \bibnamefont {Parpia}},
  \bibinfo {author} {\bibfnamefont {H.~G.}\ \bibnamefont {Craighead}}, \ and\
  \bibinfo {author} {\bibfnamefont {P.~L.}\ \bibnamefont {McEuen}},\ }\href
  {\doibase 10.1021/nl801457b} {\bibfield  {journal} {\bibinfo  {journal} {Nano
  Letters}\ }\textbf {\bibinfo {volume} {8}},\ \bibinfo {pages} {2458}
  (\bibinfo {year} {2008})}\BibitemShut {NoStop}%
\bibitem [{\citenamefont {Clougherty}(2014)}]{DPC2013}%
  \BibitemOpen
  \bibfield  {author} {\bibinfo {author} {\bibfnamefont {D.~P.}\ \bibnamefont
  {Clougherty}},\ }\href {\doibase 10.1103/PhysRevB.90.245412} {\bibfield
  {journal} {\bibinfo  {journal} {Phys. Rev. B}\ }\textbf {\bibinfo {volume}
  {90}},\ \bibinfo {pages} {245412} (\bibinfo {year} {2014})}\BibitemShut
  {NoStop}%
\bibitem [{\citenamefont {Clougherty}\ and\ \citenamefont
  {Zhang}(2012)}]{dpc11}%
  \BibitemOpen
  \bibfield  {author} {\bibinfo {author} {\bibfnamefont {D.~P.}\ \bibnamefont
  {Clougherty}}\ and\ \bibinfo {author} {\bibfnamefont {Y.}~\bibnamefont
  {Zhang}},\ }\href {\doibase 10.1103/PhysRevLett.109.120401} {\bibfield
  {journal} {\bibinfo  {journal} {Phys. Rev. Lett.}\ }\textbf {\bibinfo
  {volume} {109}},\ \bibinfo {pages} {120401} (\bibinfo {year}
  {2012})}\BibitemShut {NoStop}%
\bibitem [{\citenamefont {Clougherty}(2017)}]{dpcSPR2017}%
  \BibitemOpen
  \bibfield  {author} {\bibinfo {author} {\bibfnamefont {D.~P.}\ \bibnamefont
  {Clougherty}},\ }\href {\doibase 10.1103/PhysRevB.96.235404} {\bibfield
  {journal} {\bibinfo  {journal} {Phys. Rev. B}\ }\textbf {\bibinfo {volume}
  {96}},\ \bibinfo {pages} {235404} (\bibinfo {year} {2017})}\BibitemShut
  {NoStop}%
\bibitem [{\citenamefont {Sengupta}\ and\ \citenamefont
  {Clougherty}(2018)}]{Sengupta2018}%
  \BibitemOpen
  \bibfield  {author} {\bibinfo {author} {\bibfnamefont {S.}~\bibnamefont
  {Sengupta}}\ and\ \bibinfo {author} {\bibfnamefont {D.~P.}\ \bibnamefont
  {Clougherty}},\ }\href {\doibase 10.1088/1742-6596/1148/1/012007} {\bibfield
  {journal} {\bibinfo  {journal} {J. Phys.: Conf. Ser.}\ }\textbf {\bibinfo
  {volume} {1148}},\ \bibinfo {pages} {012007} (\bibinfo {year}
  {2018})}\BibitemShut {NoStop}%
\bibitem [{\citenamefont {Viennot}\ \emph {et~al.}(2018)\citenamefont
  {Viennot}, \citenamefont {Ma},\ and\ \citenamefont {Lehnert}}]{lehnert2018}%
  \BibitemOpen
  \bibfield  {author} {\bibinfo {author} {\bibfnamefont {J.~J.}\ \bibnamefont
  {Viennot}}, \bibinfo {author} {\bibfnamefont {X.}~\bibnamefont {Ma}}, \ and\
  \bibinfo {author} {\bibfnamefont {K.~W.}\ \bibnamefont {Lehnert}},\
  }\href@noop {} {\bibfield  {journal} {\bibinfo  {journal} {Phys. Rev. Lett.}\
  }\textbf {\bibinfo {volume} {121}},\ \bibinfo {pages} {183601} (\bibinfo
  {year} {2018})}\BibitemShut {NoStop}%
\bibitem [{\citenamefont {Chu}\ \emph {et~al.}(2018)\citenamefont {Chu},
  \citenamefont {Kharel}, \citenamefont {Yoon}, \citenamefont {Frunzio},
  \citenamefont {Rakich},\ and\ \citenamefont {Schoelkopf}}]{Schoelkopf2018}%
  \BibitemOpen
  \bibfield  {author} {\bibinfo {author} {\bibfnamefont {Y.}~\bibnamefont
  {Chu}}, \bibinfo {author} {\bibfnamefont {P.}~\bibnamefont {Kharel}},
  \bibinfo {author} {\bibfnamefont {T.}~\bibnamefont {Yoon}}, \bibinfo {author}
  {\bibfnamefont {L.}~\bibnamefont {Frunzio}}, \bibinfo {author} {\bibfnamefont
  {P.~T.}\ \bibnamefont {Rakich}}, \ and\ \bibinfo {author} {\bibfnamefont
  {R.~J.}\ \bibnamefont {Schoelkopf}},\ }\href@noop {} {\bibfield  {journal}
  {\bibinfo  {journal} {Nature}\ }\textbf {\bibinfo {volume} {563}},\ \bibinfo
  {pages} {666} (\bibinfo {year} {2018})}\BibitemShut {NoStop}%
\bibitem [{\citenamefont {de~Andres}\ \emph {et~al.}(2012)\citenamefont
  {de~Andres}, \citenamefont {Guinea},\ and\ \citenamefont
  {Katsnelson}}]{deandres2012}%
  \BibitemOpen
  \bibfield  {author} {\bibinfo {author} {\bibfnamefont {P.~L.}\ \bibnamefont
  {de~Andres}}, \bibinfo {author} {\bibfnamefont {F.}~\bibnamefont {Guinea}}, \
  and\ \bibinfo {author} {\bibfnamefont {M.~I.}\ \bibnamefont {Katsnelson}},\
  }\href {\doibase 10.1103/PhysRevB.86.144103} {\bibfield  {journal} {\bibinfo
  {journal} {Phys. Rev. B}\ }\textbf {\bibinfo {volume} {86}},\ \bibinfo
  {pages} {144103} (\bibinfo {year} {2012})}\BibitemShut {NoStop}%
\bibitem [{\citenamefont {Barclay}\ \emph {et~al.}(2006)\citenamefont
  {Barclay}, \citenamefont {Srinivasan}, \citenamefont {Painter}, \citenamefont
  {Lev},\ and\ \citenamefont {Mabuchi}}]{microdisk2006}%
  \BibitemOpen
  \bibfield  {author} {\bibinfo {author} {\bibfnamefont {P.~E.}\ \bibnamefont
  {Barclay}}, \bibinfo {author} {\bibfnamefont {K.}~\bibnamefont {Srinivasan}},
  \bibinfo {author} {\bibfnamefont {O.}~\bibnamefont {Painter}}, \bibinfo
  {author} {\bibfnamefont {B.}~\bibnamefont {Lev}}, \ and\ \bibinfo {author}
  {\bibfnamefont {H.}~\bibnamefont {Mabuchi}},\ }\href {\doibase
  10.1063/1.2356892} {\bibfield  {journal} {\bibinfo  {journal} {App. Phys.
  Lett.}\ }\textbf {\bibinfo {volume} {89}},\ \bibinfo {pages} {131108}
  (\bibinfo {year} {2006})}\BibitemShut {NoStop}%
\bibitem [{\citenamefont {Wilson}\ \emph {et~al.}(2009)\citenamefont {Wilson},
  \citenamefont {Regal}, \citenamefont {Papp},\ and\ \citenamefont
  {Kimble}}]{regal2009}%
  \BibitemOpen
  \bibfield  {author} {\bibinfo {author} {\bibfnamefont {D.~J.}\ \bibnamefont
  {Wilson}}, \bibinfo {author} {\bibfnamefont {C.~A.}\ \bibnamefont {Regal}},
  \bibinfo {author} {\bibfnamefont {S.~B.}\ \bibnamefont {Papp}}, \ and\
  \bibinfo {author} {\bibfnamefont {H.~J.}\ \bibnamefont {Kimble}},\ }\href
  {\doibase 10.1103/PhysRevLett.103.207204} {\bibfield  {journal} {\bibinfo
  {journal} {Phys. Rev. Lett.}\ }\textbf {\bibinfo {volume} {103}},\ \bibinfo
  {pages} {207204} (\bibinfo {year} {2009})}\BibitemShut {NoStop}%
\bibitem [{\citenamefont {Yu}\ \emph {et~al.}(2012)\citenamefont {Yu},
  \citenamefont {Purdy},\ and\ \citenamefont {Regal}}]{regal2012}%
  \BibitemOpen
  \bibfield  {author} {\bibinfo {author} {\bibfnamefont {P.-L.}\ \bibnamefont
  {Yu}}, \bibinfo {author} {\bibfnamefont {T.~P.}\ \bibnamefont {Purdy}}, \
  and\ \bibinfo {author} {\bibfnamefont {C.~A.}\ \bibnamefont {Regal}},\ }\href
  {\doibase 10.1103/PhysRevLett.108.083603} {\bibfield  {journal} {\bibinfo
  {journal} {Phys. Rev. Lett.}\ }\textbf {\bibinfo {volume} {108}},\ \bibinfo
  {pages} {083603} (\bibinfo {year} {2012})}\BibitemShut {NoStop}%
\bibitem [{\citenamefont {Borrielli}\ \emph {et~al.}(2016)\citenamefont
  {Borrielli}, \citenamefont {Marconi}, \citenamefont {Marin}, \citenamefont
  {Marino}, \citenamefont {Morana}, \citenamefont {Pandraud}, \citenamefont
  {Pontin}, \citenamefont {Prodi}, \citenamefont {Sarro}, \citenamefont
  {Serra},\ and\ \citenamefont {Bonaldi}}]{bonaldi2016}%
  \BibitemOpen
  \bibfield  {author} {\bibinfo {author} {\bibfnamefont {A.}~\bibnamefont
  {Borrielli}}, \bibinfo {author} {\bibfnamefont {L.}~\bibnamefont {Marconi}},
  \bibinfo {author} {\bibfnamefont {F.}~\bibnamefont {Marin}}, \bibinfo
  {author} {\bibfnamefont {F.}~\bibnamefont {Marino}}, \bibinfo {author}
  {\bibfnamefont {B.}~\bibnamefont {Morana}}, \bibinfo {author} {\bibfnamefont
  {G.}~\bibnamefont {Pandraud}}, \bibinfo {author} {\bibfnamefont
  {A.}~\bibnamefont {Pontin}}, \bibinfo {author} {\bibfnamefont {G.~A.}\
  \bibnamefont {Prodi}}, \bibinfo {author} {\bibfnamefont {P.~M.}\ \bibnamefont
  {Sarro}}, \bibinfo {author} {\bibfnamefont {E.}~\bibnamefont {Serra}}, \ and\
  \bibinfo {author} {\bibfnamefont {M.}~\bibnamefont {Bonaldi}},\ }\href
  {\doibase 10.1103/PhysRevB.94.121403} {\bibfield  {journal} {\bibinfo
  {journal} {Phys. Rev. B}\ }\textbf {\bibinfo {volume} {94}},\ \bibinfo
  {pages} {121403} (\bibinfo {year} {2016})}\BibitemShut {NoStop}%
\bibitem [{\citenamefont {Fink}\ \emph {et~al.}(2016)\citenamefont {Fink},
  \citenamefont {Kalaee}, \citenamefont {Pitanti}, \citenamefont {Norte},
  \citenamefont {Heinzle}, \citenamefont {Davanco}, \citenamefont
  {Srinivasan},\ and\ \citenamefont {Painter}}]{painter2016}%
  \BibitemOpen
  \bibfield  {author} {\bibinfo {author} {\bibfnamefont {J.}~\bibnamefont
  {Fink}}, \bibinfo {author} {\bibfnamefont {M.}~\bibnamefont {Kalaee}},
  \bibinfo {author} {\bibfnamefont {A.}~\bibnamefont {Pitanti}}, \bibinfo
  {author} {\bibfnamefont {R.}~\bibnamefont {Norte}}, \bibinfo {author}
  {\bibfnamefont {L.}~\bibnamefont {Heinzle}}, \bibinfo {author} {\bibfnamefont
  {M.}~\bibnamefont {Davanco}}, \bibinfo {author} {\bibfnamefont
  {K.}~\bibnamefont {Srinivasan}}, \ and\ \bibinfo {author} {\bibfnamefont
  {O.}~\bibnamefont {Painter}},\ }\href@noop {} {\bibfield  {journal} {\bibinfo
   {journal} {Nature Communications}\ }\textbf {\bibinfo {volume} {7}},\
  \bibinfo {pages} {1} (\bibinfo {year} {2016})}\BibitemShut {NoStop}%
\bibitem [{\citenamefont {Landau}\ and\ \citenamefont
  {Lifshitz}(1986)}]{landau}%
  \BibitemOpen
  \bibfield  {author} {\bibinfo {author} {\bibfnamefont {L.}~\bibnamefont
  {Landau}}\ and\ \bibinfo {author} {\bibfnamefont {E.}~\bibnamefont
  {Lifshitz}},\ }\href@noop {} {\emph {\bibinfo {title} {Theory of
  Elasticity}}},\ \bibinfo {edition} {3rd}\ ed.\ (\bibinfo  {publisher}
  {Butterworth-Heinemann},\ \bibinfo {year} {1986})\BibitemShut {NoStop}%
\bibitem [{\citenamefont {Nash}\ and\ \citenamefont {Sen}(1983)}]{topology}%
  \BibitemOpen
  \bibfield  {author} {\bibinfo {author} {\bibfnamefont {C.}~\bibnamefont
  {Nash}}\ and\ \bibinfo {author} {\bibfnamefont {S.}~\bibnamefont {Sen}},\
  }\href@noop {} {\emph {\bibinfo {title} {Topology and Geometry for
  Physicists}}}\ (\bibinfo  {publisher} {Academic Press},\ \bibinfo {year}
  {1983})\BibitemShut {NoStop}%
\bibitem [{\citenamefont {Ftouni}\ \emph {et~al.}(2013)\citenamefont {Ftouni},
  \citenamefont {Tainoff}, \citenamefont {Richard}, \citenamefont {Lulla},
  \citenamefont {Guidi}, \citenamefont {Collin},\ and\ \citenamefont
  {Bourgeois}}]{sin-debye2013}%
  \BibitemOpen
  \bibfield  {author} {\bibinfo {author} {\bibfnamefont {H.}~\bibnamefont
  {Ftouni}}, \bibinfo {author} {\bibfnamefont {D.}~\bibnamefont {Tainoff}},
  \bibinfo {author} {\bibfnamefont {J.}~\bibnamefont {Richard}}, \bibinfo
  {author} {\bibfnamefont {K.}~\bibnamefont {Lulla}}, \bibinfo {author}
  {\bibfnamefont {J.}~\bibnamefont {Guidi}}, \bibinfo {author} {\bibfnamefont
  {E.}~\bibnamefont {Collin}}, \ and\ \bibinfo {author} {\bibfnamefont
  {O.}~\bibnamefont {Bourgeois}},\ }\href {\doibase 10.1063/1.4821501}
  {\bibfield  {journal} {\bibinfo  {journal} {Review of Scientific
  Instruments}\ }\textbf {\bibinfo {volume} {84}},\ \bibinfo {pages} {094902}
  (\bibinfo {year} {2013})}\BibitemShut {NoStop}%
\end{thebibliography}%

\end{document}